\documentclass[12pt]{article}

 \usepackage{amsfonts}
 \usepackage{amssymb}
 \usepackage{graphicx} \usepackage{epstopdf}
 \newcommand{\crlb}[1]{\label{#1}\\[2pt]}
 
 \newcommand{\crld}[1]{\label{#1}}
 \newcommand{\eela}[1]{\quad\hbox{\scriptsize{#1}}\label{#1}\end{eqnarray}}
 \newcommand{\eelb}[1]{\label{#1}\end{eqnarray}}
 
 \newcommand{\newsecb}[2]{\section{#1}\label{#2}\setcounter{equation}{0}}

 \newcommand{\nolabels} {\def\eel{\eelb} \def\crl{\crlb} \def\newsecl{\newsecb}\def\bibiteml{\bibitem}\def\citel{\cite}\def\labell{\crld}}

\newcommand\publishversion{\nolabels\setlength{\textheight}{8.75in}\setlength{\oddsidemargin}{0in}
    \setlength{\textwidth}{6.3in}\setlength{\topmargin}{-0.3in}}

                 \def\fn{\footnote}
     \def\nm{\nonumber}   \def\be{\begin{eqnarray}}    \def\ee{\end{eqnarray}}
 \def\bi#1{\begin{itemize}\item[#1]}   \def\itm#1{\item[#1]}  \def\ei{\end{itemize}}  \def\eqn#1{(\ref{#1})}
 \def\tl#1{\tilde{#1}}  \def\^#1{\hat{#1}}

 \def\a{\alpha}      \def\b{\beta}         
 \def\d{\delta}        \def\e{\varepsilon} 
       \def\l{\lambda} \def\L{\Lambda}     \def\m{\mu}
 \def\f{\phi}            \def\vv{\varphi}    \def\n{\nu}
 \def\j{\psi}                   
         \def\th{\theta}  
               
 \def\w{\omega}      \def\W{\Omega}  

    \def\LL{{\mathcal L}}   
 \def\pa{\partial} \def\ra{\rightarrow}
  
 \def\dd{{\rm d}}

 \def\fract#1#2{{\textstyle{#1\over#2}}}
 \def\ffract#1#2{\raise .2 em\hbox{$\scriptstyle#1$}\kern-.3em/
                 \kern-.2em\lower .15 em \hbox{$\scriptstyle#2$}}
 
 \def\half{\fract12} \def\quart{\fract14}

\def\bmatrix{\begin{matrix}} \def\ematrix{\end{matrix}} \def\bpmatrix{\begin{pmatrix}}\def\epmatrix{\end{pmatrix}}
\def\bcenter{\begin{center}} \def\ecenter{\end{center}}

\def\lowerheightfig#1#2#3{\(\raise-#1\hbox{\includegraphics[height=#2]{#3}}\)}
\def\lowerwidthfig#1#2#3{\(\raise-#1\hbox{\includegraphics[width=#2]{#3}}\)}

\publishversion
\begin{document} \begin{titlepage}

\title{{ \LARGE\bf Singularities, horizons, firewalls, and local conformal symmetry} \\[10pt]
\large Karl Schwarzschild Memorial Lecture} 
\author{Gerard 't~Hooft}
\date{\normalsize Institute for Theoretical Physics \\
Utrecht University  \\[10pt]
 Postbox 80.195 \\ 3508 TD Utrecht, the Netherlands  \\[10pt]
e-mail:  g.thooft@uu.nl \\ internet: 
http://www.staff.science.uu.nl/\~{}hooft101/\vskip10pt}

\maketitle

\begin{quotation} \noindent {\large\bf Abstract } \medskip \\
The Einstein-Hilbert theory of gravity can be rephrased by focusing on local conformal symmetry 
as an exact, but spontaneously broken symmetry of nature. The conformal component of the metric field
is then treated as a dilaton field with only renormalizable interactions. This imposes
constraints on the theory, which can also be viewed as
 demanding regularity of the action as the dilaton field variable tends to 0. In other words, we have constraints
on the small distance behaviour. Our procedure appears to turn a black hole into a regular, topologically trivial soliton
without singularities, horizons of firewalls, but many questions remain.
\end{quotation}
\vfill\vfill
{\footnotesize{Presented at the 2nd Karl Schwarzschild Meeting on Gravitational Physics, Frankfurt, July 23, 2015.}}
\vfill \flushleft{\today}

\end{titlepage}

\eject
\setcounter{page}{2} 

\newsecl{Introduction}{Intro}
The modern representation of Karl Schwarzschild's spherically symmetric solution of Einstein's equations reads\fn{In Schwarzschild's original work\,\cite{KS}, the coordinate \(r\) in Eq.~\eqn{Schw} was called \(R\), while he chose an other radial coordinate \(r\) such that the point \(R=2M\) corresponds to \(r=0\), since it seemed to be obvious to expect a singular mass distribution at the origin of the coordinate frame. Today, we know that this was unnecessary, for two reasons: first, one is free to choose the most convenient coordinate system anyway, and secondly, the surface \(r=2M\) does not represent a physical singularity at all, but just a coordinate singularity, much like the north pole of the Earth. It is the black hole horizon.} 
	\be \dd s^2=-\left(1-\fract{2M}r  \right)\dd t^2+{1\over 1-2M/r}\dd r^2+r^2(\dd\th^2+\sin^2\th\,\dd\vv^2)\ . \eel{Schw}
As we now know very well, matter can enter the black hole through the horizon, defined by the surface \(r=2M\), while in the standard, unquantised theory, nothing can emerge out of it. The horizon is a one way door.\fn{On some web pages, these facts are still being disputed, which we can only attribute to ignorance. Schwarzschild, who wrote his paper in less than two months after Einstein's discovery, could be excused for not immediately realising the rather subtle features of black hole horizons, which required several years to be cleared up, but today's experts cannot afford to make such mistakes.} 
In the coordinates of Eq.~\eqn{Schw}, the point \(r=0\) is a real physical singularity.

Even though the horizon appears to be a regular region of space-time, we do have a problem with it. According to Hawking's well-known result\,\cite{SWH}, it is due to vacuum fluctuations that a distant observer will observe particles leaving the black hole: Hawking radiation.
These particles have a thermal spectrum, independent of the black hole formation process. 

Hawking's original conclusion was that this result must imply that a black hole as a physical object violates the laws of quantum mechanics: even if it originates from matter in a single quantum state, it ends up in a thermal, that is, a quantum mechanically mixed state. How could it be that a derivation that uses quantum mechanics can yield a result violating the laws of this theory? Hawking particles are now understood to be formed at the horizon, not, as was originally thought, somewhere near the \(r=0\) singularity in its past.

According to the present author's understanding of quantum mechanics\,\cite{GtHqm}, however, all states in which the Hawking particles fluctuate differently, are different ontological states of the system, and they should be treated as different quantum states as well. Thus, the vacuum state, which is a single quantum state, emerges at the horizon as a collection (superposition) of infinitely many ontological states, and it should be treated as such. One can then understand how particles entering a black hole, can affect these ontological states in spite of the fact that their probabilistic distribution remains unaltered. This effect can actually be calculated\,\cite{GtHgravshift}.

In Ref.~\cite{GtHqm}, version 3, it is explained how pure states can transform into mixed states under general coordinate transformations.

\newsecl{Local conformal symmetry}{locconf}

It is to be noted that the main features that went into our description of the back reaction to Hawking radiation only requires knowledge of light-like geodesics. These depend on all components of the metric tensor \(g_{\m\n}(x)\), except for one overall factor. This is because the equation for light-like geodesics,
	\be \dd s^2=g_{\m\n}\dd x^\m\dd x^\n=0\ , \eel{lightgeo}
is unaltered by the substitution
	 \be g_{\m\n}(x)\ra \W^2(x)\,g_{\m\n}(x)\eel{conftrf}
(take into consideration that this equation alone, without higher derivatives, determines the shapes of all light cones). Theories invariant under \eqn{conftrf} are said to be locally conformally invariant. By adding a dilaton field, as will be explained shortly, even the Einstein-Hilbert action can be made invariant under \eqn{conftrf}, because the entire metric tensor, including its common factor, consists of dynamical variables. \(\sqrt {-g}\) is not invariant, but covariant. This implies that flat space time, that is, the vacuum state, breaks the symmetry. Thus we say that conformal symmetry is not explicitly, but spontaneously broken in Enstein-Hilbert gravity, just as local \(SU(2)\times U(1)\) gauge symmetry is spontaneously broken by the BEH mechanism.

We write the standard Lagrangian for gravity interacting with matter as \def\EH{{\mathrm{EM}}} \def\mat{{\mathrm{matter}}}
	\be \LL&=& \LL^\EH\ +\ \LL^\mat \ ;\qquad  \LL^\EH\ =\  \fract 1{16\pi G}\sqrt{-g}\,(R-2\L)\ , \\
		\LL^\mat&=& \LL^{YM}(A)+\LL^{\mathrm{bos}}(A,\,\f,\,g_{\m\n})+\LL^\mathrm{ferm}(A,\,\j,\,\f,\,g_{\m\n})\ , \eel{Lmatter}
where \(\f(x)\) represents the scalar matter fields, and \(\j(x)\) the fermionic ones. \(A\) stands for \(A_\m(x)\), the Yang-Mills fields in the matter Lagrangian. Now define	
	\be g_{\m\n}=\w^2(\vec x,t)\,\^g_{\m\n}\ ;\qquad\LL=\LL(\w,\,\^g_{\m\n},\,A_\m,\j,\f)\ .\eel{omegasplit}	
This contains the `trivial' conformal symmetry
	\be \^g_{\m\n}\ra\W^2(\vec x,t)\^g_{\m\n}\ ,\quad \w\ra\W^{-1}\w\ ,\quad A_\m\ra A_\m\ ,\nm	\\
		\f\ra\W^{-1}\f\ ,\quad\j\ra\W^{-3/2}\j\ . \eel{confsymm}
We shall refer to the field \(\w(\vec x,t)\) as the \emph{dilaton field}. 

Working out the Einstein-Hilbert Lagrangian and the matter Lagrangian a bit more explicitly gives
	\be \LL^\EH&=&\sqrt{-\^g}\left(\fract 1{16\pi G}(\w^2\,\^R+6\^g^{\m\n}\pa_\m\w\pa_\n\w)\ -\ \fract{\L}{8\pi G}\,\w^4\right)\ ; \\
		\LL^\mat &=& -\,\quart F_{\m\n}F_{\m\n} \ + \nm\\
		&&\sqrt{\^g}\left(-\half \^g^{\m\n}D_\m\f D_\n\f-\half m^2\w^2\f^2-\fract 1{12} \^R\f^2 - 
			 \fract{\l}{8}\f^4\right)+\LL^{\mathrm{ferm}}\ .\quad \ee
Here, we included the \(\^R\,\f^2\) term for restoring conformal invariance of \(\LL^\mat\), where \(\^R\) is the scalar curvature associated to \(\^g_{\m\n}\).

Now, several remarks are of order: \def\tot{{\mathrm{tot}}}  \def\kin{{\mathrm{kin}}}
\bi{-}surprisingly perhaps, the Einstein-Hilbert action appears to be an entirely renormalizable Lagrangian for the dilaton field \(\w(x)\).
\itm{-} With the cosmological term acting as a quartic coupling term, the matter Lagrangian for the scalar field \(\f(x)\) has  the same form as \(\LL^\EH\), apart from a factor \(-4\pi G/3\). 
\itm{-} This factor can easily be taken care of by rescaling the \(\w\) field, but its sign is curious. Since it so happens that the Standard Model neither contains explicit mass terms for the fermions, nor cubic couplings among the saclar fields\fn{Such  terms would come with a factor \(i\w\), and hence appear to violate unitarity.}, we can include a factor \(i\) in the redefinition of \(\w\) without any obvious violation of unitarity.
\itm{-} Due to the necessary rescaling of \(\w\), all physical constants (including mass terms and the cosmological term) eventually emerge as dimensionless combinations of Newton's constant \(G\) and the Standard Model parameters. \ei
Nevertheless, the theory is non renormalizable. This is because a kinetic term for the \(\^g_{\m\n}\) field is missing. Normally, theories cease to be renormalizable if a kinetic term is missing. The theory would be ill-defined altogether, but by inspecting the way one would normally handle the Einstein equations in perturbation expansions, one finds the following formal prescription for solving the classical equations:
\begin{quote} Find the total energy-momentum-stress tensor \(T_{\m\n}^\tot\) for the matter fields, including the \(\w\) field. Note that the original, \(\w\)-independent Einstein-Hilbert action disappeared, so that Einstein's equation is to be replaced by one where Newton's constant is infinite. Therefore, the equations are:
	\be T_{\m\n}^\tot\ =\ T_{\m\n}^\mat-T_{\m\n}(\w)\ =\ \ 0\ \ =\ T_{\m\n}^\mat-{\textstyle{\frac 1{8\pi G}}}\,G_{\m\n}\ . \eel{totem} \end{quote} 
We kept the minus sign in the contribution of the \(\w\) field; it disappears when the factor \(i\) mentioned above is employed. We recognise Einstein's original equation of course; however, in the conformally symmetric notation, we should say that the condition that the total energy-momentum-stress tensor vanishes is a constraint. It has exactly the right dimension to enforce the equations for the \(\^g_{\m\n}\) field (the conformal stress-energy momentum tensor is traceless).

If our aim were to restore renormalizability, all we had to do now would be to collect all divergent diagrams and determine their general form. This should provide us with terms to be added to the original bare Lagrangian of the theory, as is usually done. In this case, we find that all  divergent expressions unaccounted for, contain external lines for \(\^g_{\m\n}\) and factors \(k^4\) because they are quartically divergent. Since all calculations should be performed while respecting local conformal invariance, and all diagrams are polynomials in the fields, one expects that the only terms that should be added in the Lagrangian are locally conformally invariant expressions with four derivatives in the metric fields \(\^g_{\m\n}\). There exists only one such term that is invariant under local conformal transformations, the Weyl action:
	\be \LL^\kin=-\fract{\l^W} {2}\,C_{\m\n\a\b}\,C_{\m\n\a\b}\quad\ra\quad-
			\fract{\l^W}{4}(\pa^2 \^g_{\m\n}^{\mathrm{transverse}})^2\ . \eel{Weyl}
With this term added, the theory indeed becomes renormalizable, as is well-known, but there appear to be two complications: first, the Weyl term would generate propagators for \(\^g_{\m\n}\) that are quartic in the momenta. This is not in accordance with standard prescriptions in renormalization theory. Propagators ought to be quadratic in the momenta, in a carefully prescribed way, in order to comply with unitarity, causality, and positivity of the energy. Does this mean that our theory is not unitary, or is its energy not bounded from below? Note that the energy momentum tensor is required to obey  Eq.~\eqn{totem}, so that the total energy vanishes strictly, but that was before we added the Weyl action. What is the unitarity/energy condition in the case of conformal invariance?

Secondly, there is an other mystery. When the required renormalization counter term is computed without keeping track of conformal symmetry, one finds\,\cite{Duff} that it does not take the form \eqn{Weyl}, since also \(\sqrt{-g}\,R^2\) terms appear. Now, if we do use the conformal notation,  this would generate \(\pa_\m\w/\w\) and \(\pa_\m\w^2/\w^2\) terms, which of course cannot come from symmetric diagrams. The exact cause of this discrepancy is not yet quite understood, but may have to be interpreted as an anomaly.

Indeed, when inspecting the scaling behaviour of this theory, one finds all those anomalies that generate the renormalization group \(\b\) coefficients.
In gauge theories, we are accustomed to imposing the requirement that anomalies that destroy gauge invariance must be arranged to cancel out. We expect the same in this theory: \emph{all conformal anomalies must be demanded to cancel out.} This means that all renormalization group \(\b\) coefficients must be demanded to vanish. Consequently, all coupling parameters must be adjusted such that they are at a zero of their \(\b\) functions. This generates at least as many constraints as there are coupling parameters of the theory. We are lead to an exciting speculation: In gravity theories with conformal invariance, \emph{all} physical constants, including the masses and even the cosmological constant, are constrained to values that in principle must be computable. 

In short, we propose that conformal symmetry is not just an accident that vaguely applies to some branches of physics, but that it may play a very important role as an absolutely exact transformation rule. It will then be  an essential instrument that might lead us towards calculating 
parameters that otherwise would have been freely adjustable, and a crucial ingredient of the description of black holes, as we will see.

\newsecl{Black holes}{BH}

In a nut shell, the \emph{black hole information problem} is  the question how information concerning the state of matter entering the hole, can be seen to be present in the particles coming out, as was mentioned in the Introduction. One way of phrasing this difficulty is the question how to avoid that the information entering the hole disappears into the central singularity. A related difficulty arises if one considers the entanglement of particles entering the hole and others that emerge. The contradictions appear to be strong enough to make some researchers\,\cite{firewall} believe that a \emph{firewall} should emerge at the horizon, prohibiting particles to enter.
	
	Conformal symmetry will be of help here: the central singularity disappears\fn{This happens as follows: near the singularity, we can stretch the coordinates  so much, that the curvature-squared will no longer be singular, but instead, the singularity moves to the infinite future. In a sense, that is where it belongs anyway. A good exercise is to multiply the metric with an overall factor such as \(1/r^4\), to see how this makes the singularity move towards the infinite future, where space-time becomes locally flat.}, and the horizon will become ``fuzzy".	Consider the mass \(M\) of the black hole. An observer \(A\) falling in passes the horizon, experiencing the metric \(g_{\m\n}\) associated to the mass \(M\). An outside observer \(B\), however, may observe the Hawking radiation that causes the mass to shrink. While the ingoing observer still hovers over the horizon, seeing a fixed mass \(M\), the outside observer sees the mass shrink to zero. Who is right?

\begin{figure}[h] \setcounter{figure}{0}
 \begin{center}   
\quad\\[10pt]
\lowerwidthfig{0pt}{300pt}{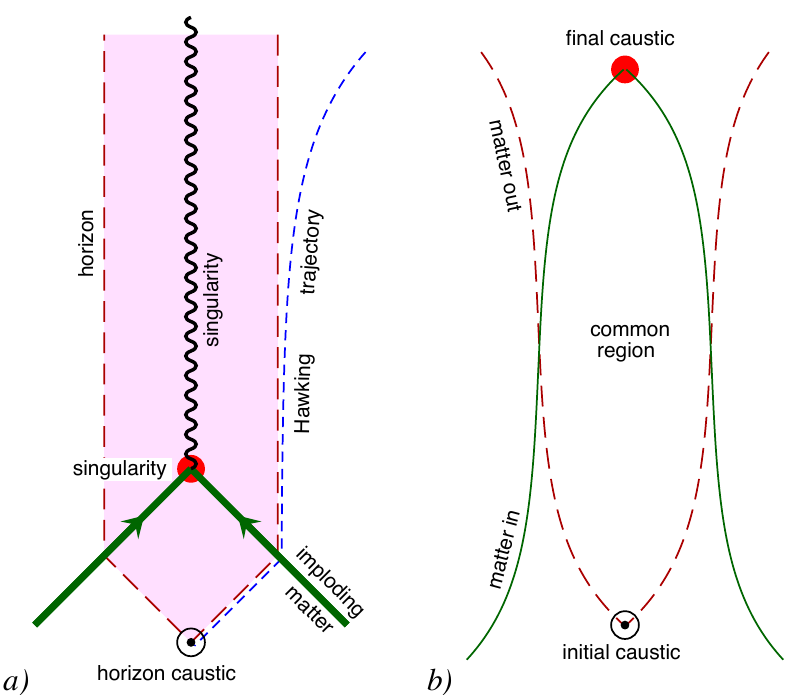}\quad  {\small{\textit{c})}}\lowerwidthfig{-20pt}{120pt}{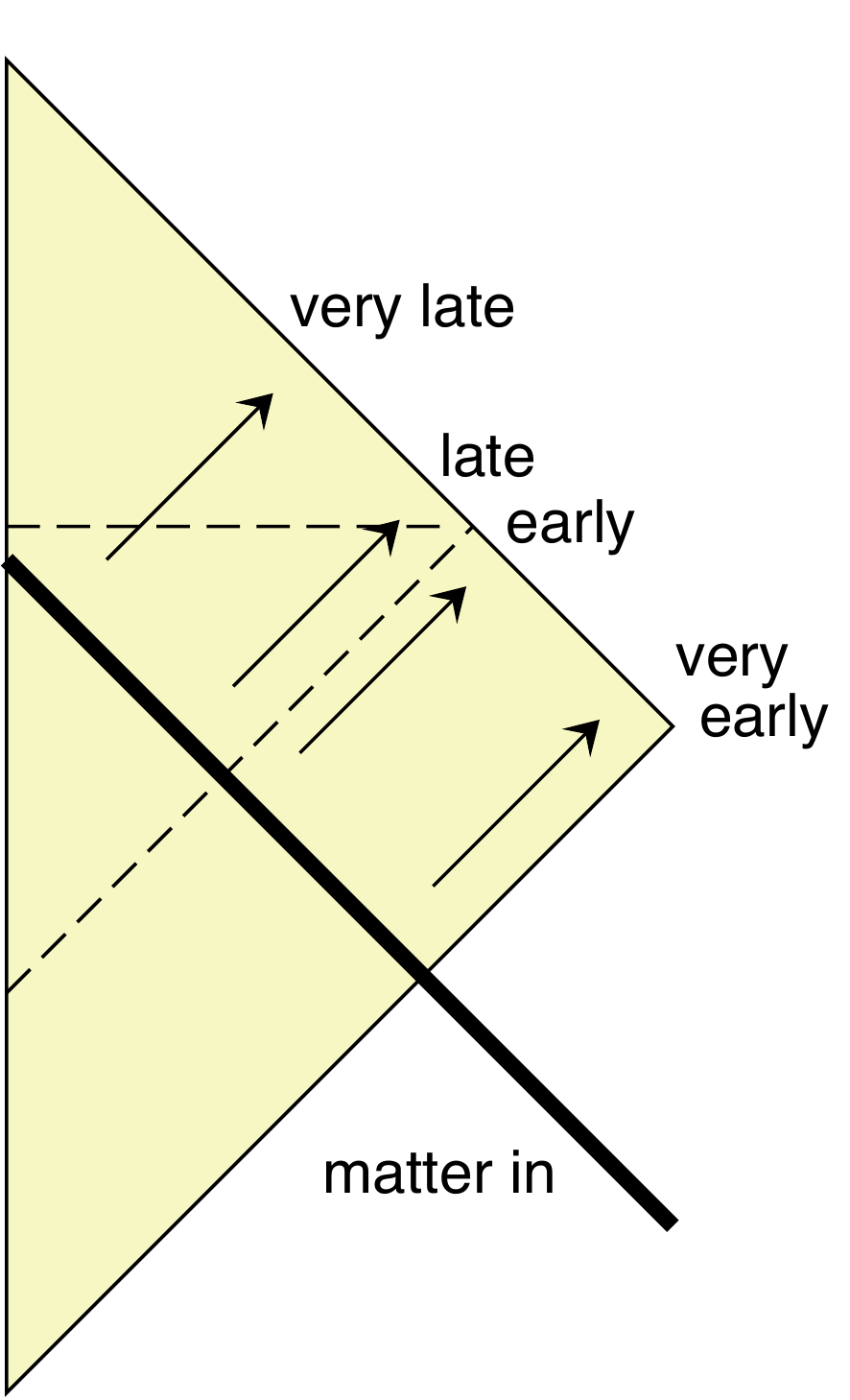}
\caption{\small $a$) Standard view of space-time of a black hole formed by imploding matter; $b$) Conformal distortion of $a$, such that time reversal symmetry is restored. $c$) Penrose diagram of symmetric black hole showing entangled states of Hawking radiation. }
\labell{softhole.fig}
\end{center} \vskip-12pt
\end{figure}

The answer may be \emph{Black hole complementarity}\,\cite{complementarity}: both observers are right, but they should use the metric \(\^g_{\m\n}\), and its conformal factor depends on who is looking. Both observers may describe their metric as
	\be \dd \^s^2=M^2(\tl t)\left(-\dd t^2 (1-{\fract 2r})+{\dd r^2\over 1-2/r}+r^2(\dd \th^2+\sin^2\th\dd\vv^2)\right)\ . \eel{varyingM}
Here, \(M(\tl t)\) may depend on the retarded time \(\tl t\), and be different for the different observers.  For the observer \(A\) entering the hole, \(M(\tl t)=M\) is constant, but for the outside observer \(B\), it goes to zero. It may also depend on the advanced time. Since the black hole has a finite life time, there is, strictly speaking, no horizon.\fn{But keep in mind that, for most of all practical purposes, there still is a horizon, as its `fuzziness' is almost imperceptible. Only for black holes close to the Planck size, this fussiness becomes important.}

\begin{figure}[h]
 \begin{center}   
\lowerheightfig{20pt}{140pt}{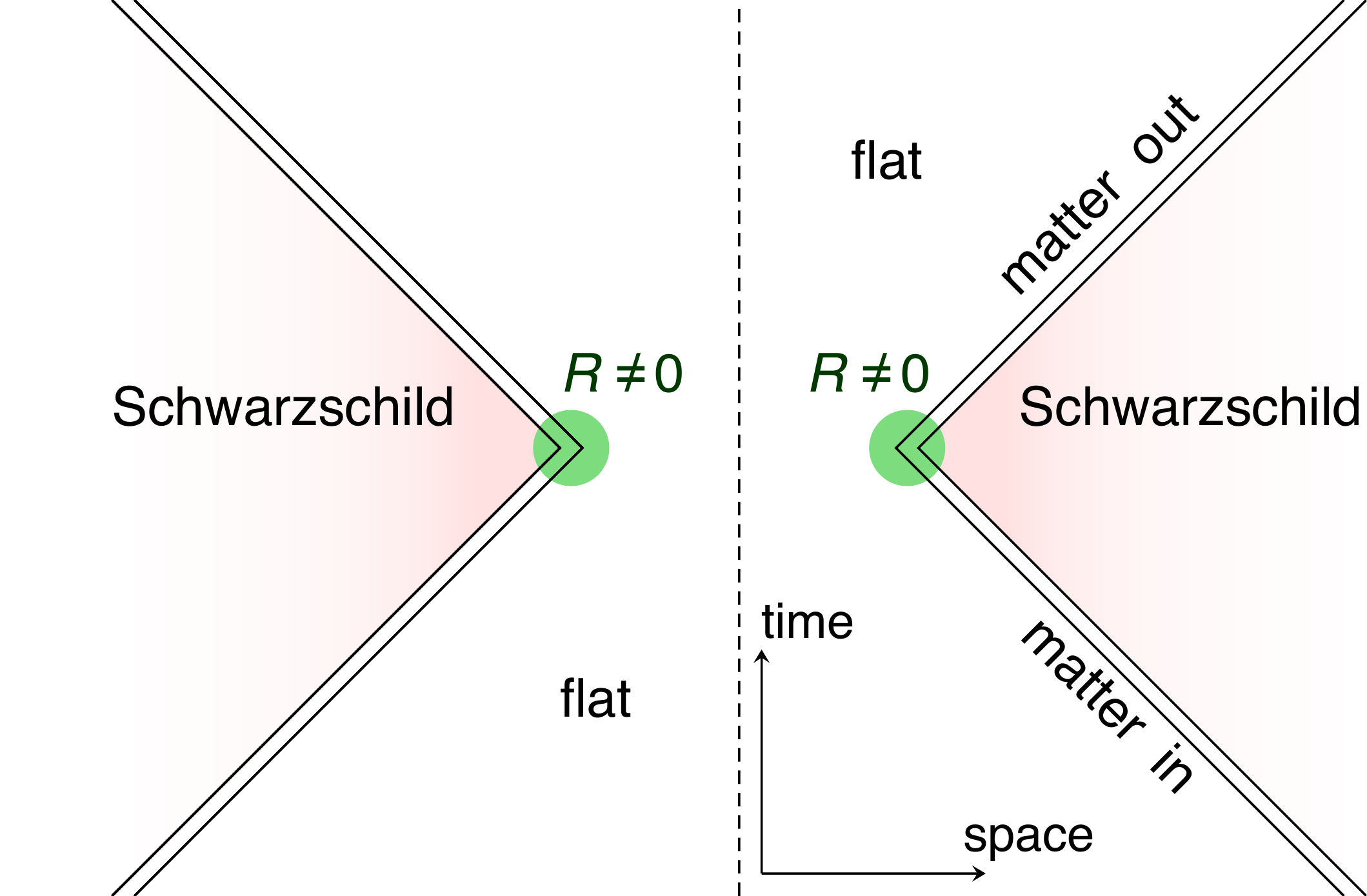}
\caption{\small The black hole metric as seen by a distant observer \(B\), see text.}
\labell{inoutmetric.fig}
\end{center} \vskip-5pt
\end{figure}
	
With `black hole complementarity',   the distant observer \(B\) sees matter going in and matter going out, The observer \(A\), going in, sees the locally clear horizon, while she cannot detect the Hawking particles. Observer \(B\) sees that the mass \(M\) vanishes during the final explosion. For this observer, the horizon produces matter, as if the imploding matter contained some sort of dynamite, causing an explosion at exactly the right moment. This observer sees an almost singular concentration of Ricci scalar curvature\fn{Note that we now use the metric \(\^g_{\m\n}\) throughout; the dilaton \(\w\) is just an `ordinary' renormalizable field, that happens to hover around its vacuum value.} at the horizon, see Fig.~\ref{inoutmetric.fig}. The curvature is strong where the future event horizon meets the past horizon. One could call this a `firewall', but the firewall is invisible for the ingoing observer \(A\). The observers use different ways to fix the `conformal gauge'. Hence, they also have different perceptions of the energy-momentum tensor of the matter present. They do both agree what the vacuum expectation value of the \(\w\) field should be, but they do not agree about what the vacuum state is. Note that, this disagreement about the vacuum state has always been a standard concept in the derivation of Hawking radiation\,\cite{SWH}\cite{GtHgravshift}.

The outside observer \(B\) sees Hawking particles emerge from a highly concentrated `curtain' along the past event horizon. If he computes the corresponding metric, he will disagree with observer \(A\), by finding an extra conformal factor. The Hawking particles cause a sharp jump in the gradients of this conformal factor. Consequently, observers \(A\) and \(B\) disagree about this conformal factor when discussing the interior region of the black hole, see Fig.~\ref{firewall.fig}.

The comparison between our spontaneously broken local conformal symmetry and the Brout-Englert-Higgs mechanism gives striking similarities. For one, this mechanism improves considerably the convergence features of the theory in the far ultra-violet. If the Weyl term \eqn{Weyl} would really be allowed then we would indeed have a renormalizable theory of gravity, as is well known. The other similarity concerns the topologically non trivial soliton solutions: before invoking the BEH mechanism, Maxwell's theory cannot allow for a singularity-free description of magnetic monopoles, while in some versions of the BEH models one can have regular monopoles; similarly, with local conformal symmetry, black holes can be made singularity free.

Our demand that all conformal anomalies cancel is a severe constraint on the theory, but this may actually be a welcome feature; it may imply that constants that are not normally computable may now be found to obey (interesting) equations. A difficulty here is that all couplings may emerge as being large, in which case we cannot perform perturbative calculations. By carefully choosing the algebra, depending on a large integer \(N\), one sometimes can search for solutions with couplings proportional to \(1/N\) or \(1/N^2\), allowing us to do \(1/N\) expansions. A very preliminary search was only partially successful, as it did allow for \(1/N\) expansions, but it did not lead to physically interesting solutions.
\begin{figure}[h]
 \begin{center}   
\lowerheightfig{20pt}{120pt}{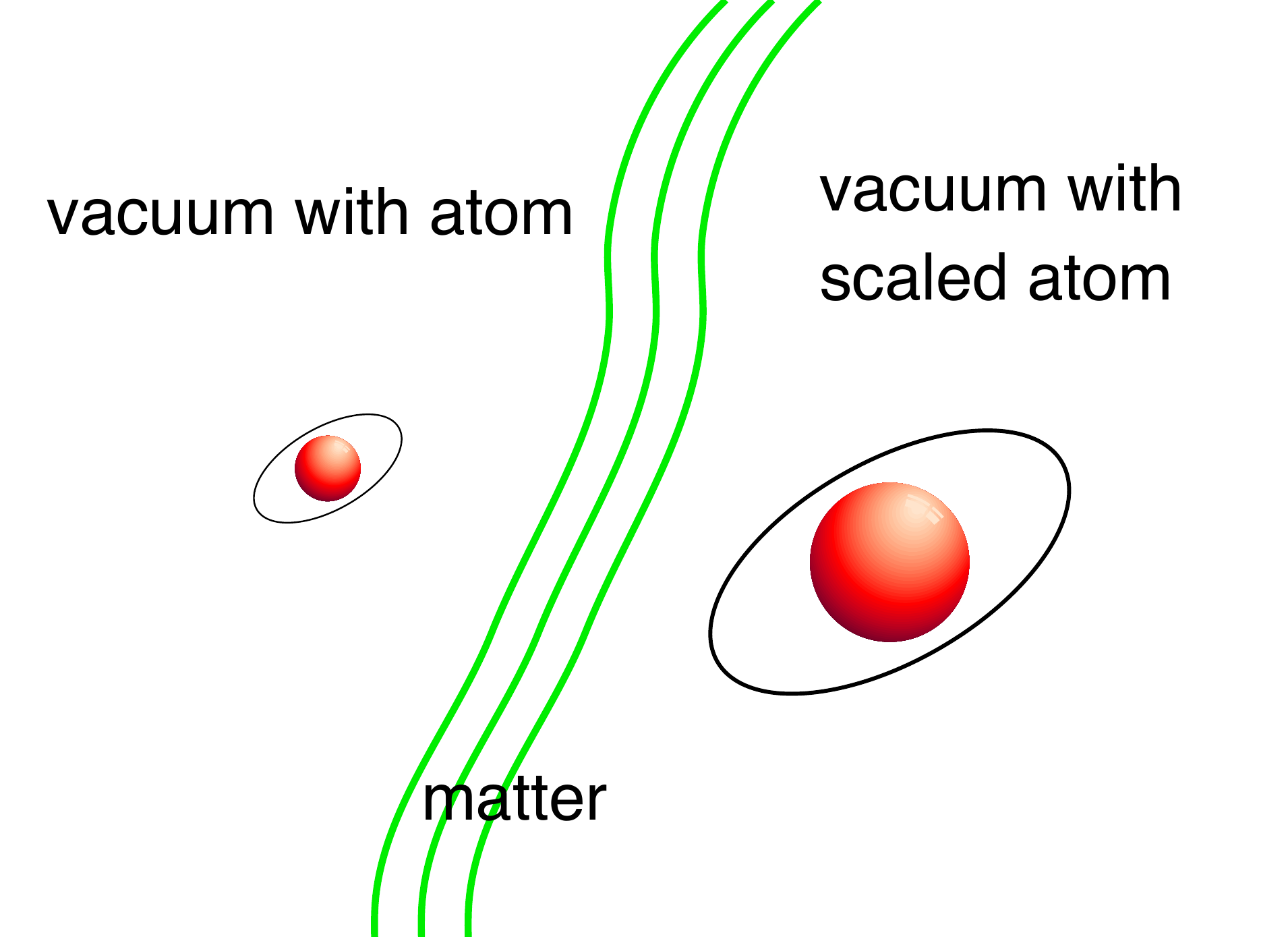}
\caption{Observer \(B\) experiences a firewall, formed by Hawking particles, while \(A\) sees no such thing. Therefore, they disagree about sizes of things inside the black hole, see text.}
\labell{firewall.fig}
\end{center}
\end{figure}

\newsecl{Features and limitations}{limit}

We do observe that spontaneously broken local conformal symmetry holds the promise that the unknown parameters of the matter Lagrangian, today described by the Standard Model, are not freely adjustable but can be computed. The numbers, however, will depend on the algebra of the matter theory. For sure, the algebra of the Standard Model will need considerable extensions in order for it to be applicable all the way to the Planck scale.

The most fundamental obstacle was found to be the \emph{hierarchy problem}: ratios of physical constants in the real world contain very large or small numbers such as \(10^{-122}\) for the cosmological constant, whereas in principle our models should turn up numbers of order 1 in Planck units. Our universe owes its complexity to the existence of exotic large numbers. One might bring this forward as an objection to our theory but it has to be remembered that the hierarchy problem is the source of headaches for many other theories as well. Barring the ``anthropic principle", no theory is known that can account for the complexity of our universe.

An other mystery is the apparent lack of unitarity. This feature was further investigated. If one chooses the coefficient \(\l^W\) in the Weyl term \eqn{Weyl} large, the theory allows for a perturbative analysis. If we add this extra term to the original Einstein Hilbert action (in the old, non conformal notation), one finds that \(\l^W\) has the dimensionality of an inverse mass squared, 
	\be \l^W\equiv 1/M^2\ , \eel{Weylmass}
where \(M\) locates poles in the complex momentum plane, and it is small compared to the Planck mass (in a renormalizable theory, a Planck mass large compared to the mass scale \(M\) of the theory indicates small gravitational couplings, hence the usefulness of perturbative expansions).

We then look at plane waves of the theory, finding that the wave equations indeed contain new poles. Identifying the quantum numbers of these poles, we found:
\bi {-} one massless pole, describing the familiar graviton. This was to be expected because the far infrared region should not be affected by the Weyl term, as it contains extra derivatives. The graviton has spin 2, but, being massless, has only two physical helicities, as in the usual theory;
\itm{-}  poles of the form \(1/(k^2+M^2-i\e)\). They all turned out to be at the same mass value \(M\). One pole has helicities \(\pm 2\), one has helicities \(\pm 1\) and one pole has helicity 0. We recognise this as the five helicities of a single, massive spin 2 particle. The problem with these poles is that they all have the wrong overall sign in the propagator.
The fact that this wrong sign is inevitable can easily be seen from the far ultraviolet limit. There, only the Weyl term contributes to the graviton propagator. It was inevitable that we have there:
\be 1/(k^2+\l^Wk^4)=1/k^2-1/(k^2+M^2)\ . \eel{cancelpoles}
\itm{-}The scalar pole that might be generated by the conformal factor, as usual in perturbative gravity, is a ghost, so that it can be ignored, it is not a physical particle, while the others seem to be real. \ei
Thus we have a single, negative metric, spin 2 companion of the graviton, with 5 possible helicity states. We propose the name ``gravitello" for that, the mysterious companion of the graviton. Having such a particle seems to be inevitable. 

We do not know how to accommodate for it in a unitary theory, but one could consider the following notions.

Our fields have oscillation modes with opposite signs. The energies can be written as
	\be H=|\vec k|(p_1^2+x_1^2)-\sqrt{\vec k^2+m^2}\,(p_2^2+x_2^2)\ , \eel{harmopposite}
where \(\vec k\) is the spacelike momentum of the wave, while \(x_i\) are the fields and \(p_i\) are the canonical momenta of these fields, at wave number \(\vec k\). They obey the usual commutation rules 
	\be [x_i,\,x_j]=[p_i,\,p_j]=0\ ,\qquad[x_i,\,p_j]=i\,\d_{ij}\ . \eel{xpcomm}
We can write Eq.~\eqn{harmopposite} as
	\be H=A\,a^\dag a-B\,b^\dag b+C\ ,\eel{creation}
where \(a\) and \(a^\dag\) are the annihilation and creation operators of the graviton, and \(b,\ b^\dag\) those of the gravitello, both having momentum \(\vec k\).

One approach is the following. We could impose a lower bound to the energies of the modes with the wrong sign, by putting a limit on the occupation numbers of the \(b,\ b^\dag\) operators. This requires the interchange  \(b\leftrightarrow b^\dag\), which is possible if
 we can rearrange and renormalise the quantum states reached by these operators. Effectively, this switches the sign of the commutator of \(b\) and \(b^\dag\). The associated replacement in the operators \(x_2\) and \(p_2\) implies that they commute as purely imaginary fields would:
\be x_2\ra ix_2\ ,\qquad p\ra ip\ ,\eel{imxp}
that is, the field of the gravitello should be chosen to be purely imaginary. This is the same operation as was required in the quantisation of ordinary gravitation; there also, the overall conformal factor, which would contribute with the wrong sign to the Einstein Hilbert action, must be replaced by a purely imaginary Lagrange multiplier field.

The procedure described here is related to the author's proposals for the interpretation of quantum mechanics\,\cite{GtHqm}. All harmonic oscillators that we encounter in the physical world, should be associated with processes in an ontological underlying world that are \emph{periodic} in time. When harmonic oscillators interact, they cease to be exactly periodic, and this means that, also in the ontological underlying world, the associated processes are no longer exactly periodic. 

To identify the quantum states of the oscillator with the classical states of the ontological world\,\cite{GtHqm}, we have to discretise them. This is achieved in `cogwheel models', which are periodic but have only a finite number of states. The annihilation and creation operators \(a\) and \(a^\dag\) are then replaced by the operators \(L_-\) and \(L_+\) in a large \(\ell\) representation of the SU(2) rotation algebra of angular momenta. These decrease or increase the quantum number \(m=L_3\), which has both a lower and an uper bound: \(|m|\le\ell\).

Regarding the gravitello, we should hasten to add that this approach has not yet been elaborated in a satisfactory way, since the gravitello would couple with an imaginary coupling constant to gravitational sources, and we have not succeeded in showing how this can be squared with unitarity.

\newsecl{Conclusion}{conc}
Local conformal symmetry shines a new light on problems where gravity couples to matter, both in the domain of the Standard Model and in our understanding of black holes. It is always our intention to make the smallest possible modifications in our models of physics, because pure Einstein-Hilbert gravity, coupling just with the Standard Model particles, appears to agree with observations extremely well, and lessons learned from past experiences suggest that one should not abandon known facts in the natural world too easily. 

It so happens that exact local conformal invariance does play a role already in Einstein-Hilbert gravity itself, by isolating the dilaton component of the metric, so the only `new' thing we add to this is the demand that this symmetry should be exact, rather than having it as an accidental, approximate feature. 

It may be noted that our approach is related to the theory of `asymptotic safety'\,\cite{safety}.  In this theory, the UV limit of gravity theory tends to a fixed point. At this fixed point, such models should also enjoy scale invariance, \emph{ergo} local conformal invariance. There, however, one is confronted with strong interactions (since the fixed point values of the coupling parameters are not close to zero). In our models, we offer perturbative accessability by adding explicit interactions (the Weyl term). This could be an advantage, but it would also introduce wrong metric states that we have to handle, somehow. In fact, whether asymptotically safe models contain states with the wrong sigm of the metric and/or the energy, is not known.

Our procedure appears to suggest that all freely adjustable parameters, being both the masses and the coupling parameters of the Standard Model, including eventually the cosmological constant, must  be computable. Our point is that, with the dilaton field  \(\w\) added, the matter component of our particle system should be completely conformally invariant, so that the physical parameters, all starting out as being dimensionless, must be exactly at the fixed point, that is, the point where all \(\b\) coefficients vanish. 

The inclusion of the Weyl interaction, which on the one hand seems to be inevitable, does generate severe problems of negative metric states, or equivalently, negative energy states. This may simply mean that we have not yet fully understood what local conformal symmetry really is\,\cite{Mannheim}.

\end{document}